# The pipeline for the ExoMars DREAMS scientific data archiving

P. Schipani,[1] L. Marty,[1] M. Mannetta,[1] F. Esposito,[1] C. Molfese,[1] A. Aboudan,[2] V. Apestigue-Palacio,[3] I. Arruego-Rodriguez,[3] C. Bettanini,[2] G. Colombatti,[2] S. Debei,[2] M. Genzer,[4] A-M. Harri,[4] E. Marchetti,[5] F. Montmessin,[6] R. Mugnuolo,[5] S. Pirrotta,[5] C. Wilson,[7]

[1] *INAF - Capodimonte Astronomical Observatory, I-80131 Naples, Italy; pietro.schipani@oacn.inaf.it*

[2] *CISAS, Via Venezia 15, I-35131 Padua, Italy*

[3] *Instituto Nacional de Tecnica Aeroespacial (INTA), Madrid, Spain*

[4] *Finnish Meteorological Institute (FMI), Erik Palmenin aukio 1, FI-00560 Helsinki, Finland*

[5] *Agenzia Spaziale Italiana (ASI), Via del Politecnico, I-00133 Roma, Italy*

[6] *LATMOS - CNRS, Quartier des Garennes, 11 bd d'Alembert, F-78280 Guyancourt, France*

[7] *Oxford University, Parks Road, Oxford, United Kingdom*

**Abstract.** DREAMS (Dust Characterisation, Risk Assessment, and Environment Analyser on the Martian Surface) is a payload accommodated on the Schiaparelli Entry and Descent Module (EDM) of ExoMars 2016, the ESA and Roscosmos mission to Mars (Esposito (2015), Bettanini et al. (2014)).

It is a meteorological station with the additional capability to perform measurements of the atmospheric electric fields close to the surface of Mars. The instrument package will make the first measurements of electric fields on Mars, providing data that will be of value in planning the second ExoMars mission in 2020, as well as possible future human missions to the red planet.

This paper describes the pipeline to convert the raw telemetries to the final data products for the archive, with associated metadata.

## 1. Mission Phases

With reference to the mission phase definitions, the DREAMS experiment operations and checkout sessions are foreseen during the Near Earth Commissioning Phase (NECP), the Interplanetary Cruise Phase (ICP) and the Primary Science Phase (PSP). During NECP and ICP, the functionality of DREAMS are tested through checkout sessions and Mission TimeLine (MTL) dumps. DREAMS scientific measurements are performed after landing on Mars, i.e. the instrument starts the science phase promptly after the touchdown. After its activation on the Martian surface, DREAMS operates according to its own pre-planned timelines.

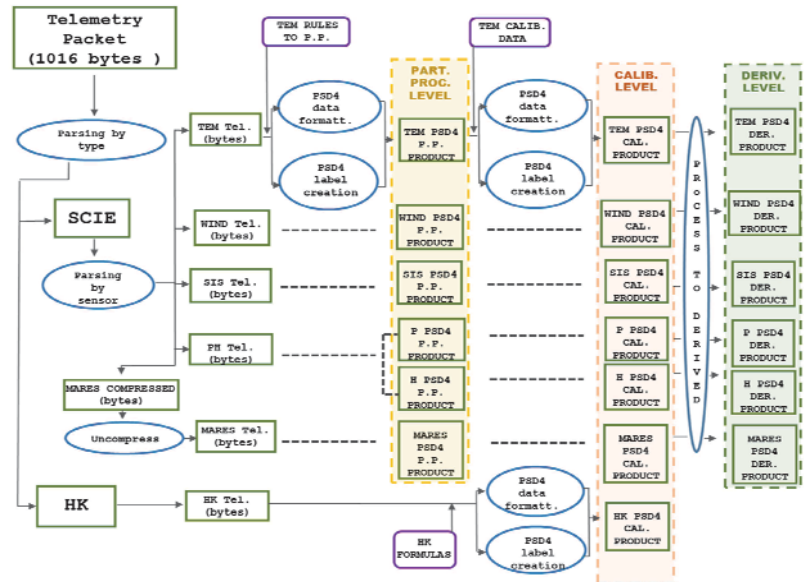

Figure 1. The DREAMS data pipeline.

## 2. Standards

The DREAMS data are archived into the European Space Agency's Planetary Science Archive (PSA), the central repository for all scientific and engineering data returned by ESA's Solar System missions. The ExoMars mission and consequently the DREAMS archive adopts the NASA's Planetary Data System (PDS) standards as a baseline for the formatting and structure of all data. The PDS standard contains requirements in terms of data set structure and documentation. Each data product is associated to a label containing full details on the structure and content of the product. The ExoMars mission and DREAMS adopt the PDS version 4 standards, acknowledged as PDS4. It has a modernized approach to archiving data within the PDS; labels are expressed as XML documents that are tied to a centralized, self-consistent model providing uniformity across the PDS.

## 3. Data Flow

The data processing pipelines, validation and delivery of the products for the ingestion into the Planetary Science Archive follow the ESA guidelines.

Immediately after the data becomes available in the Egos Data Disposition System (EDDS), telemetry packets, housekeeping parameters and any additional information relevant for data processing and analysis are retrieved.

The instrument team generates partially processed, calibrated and derived data products based on the best calibration factors and analysis routines. The data products are stored in a local data repository as well as sent to ESAC for validation and ingestion to the ESA Planetary Science Archive (PSA). Figure 1 describes the conversion process from telemetry to data products in PDS4 format, ready for upload to ESA. Using the information stored within the telemetry packet headers, first the received packets are identified as science or housekeeping data. In case of scientific packets, the next operation in the pipeline is to parse and extract from the packet payload the sample of the different sensors and to convert them to PDS4 format at the different data processing level. In case of housekeeping packets, the operations are similar but simplified.

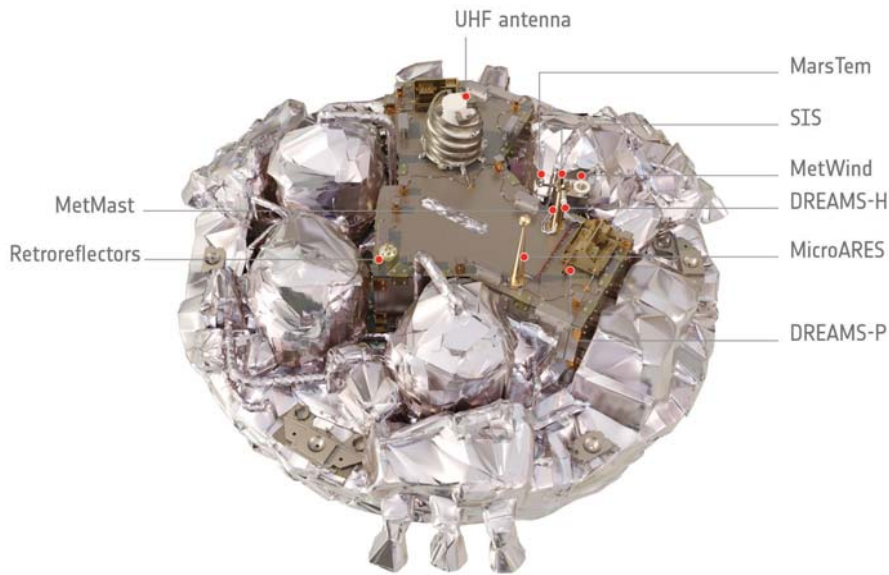

Figure 2.   The DREAMS sensors on the Schiaparelli entry, descent and landing demonstrator module (EDM). Other devices are a compact array of laser retroreflectors, attached to the zenith-facing surface of Schiaparelli (that can be used as a target for Mars orbiters to laser-locate the module) and a UHF antenna, used for communicating with the Trace Gas Orbiter. (Credit: ESA/ATG medialab).

The data transfer between DREAMS and the S/C is implemented using the CANOpen protocol: data are transferred in the form of Service Data Objects (SDOs) or Process Data Objects (PDOs). The PDOs are 8 byte data packets, adopted e.g. to transmit housekeeping information to the S/C. The SDOs are buffers of 1016 bytes, consisting of a 16 byte header field and a 1000 byte payload field, which can contain science or housekeeping data, or dump of timelines.

The generation of raw level products from telemetries implies a basic knowledge of the telemetry buffer structure. The structure of telemetries coming from DREAMS are used to develop the conversion software implemented by ESAC.

The generation of next level products implies the decoding of the telemetry buffers. The DREAMS partially processed data are values converted from ADC units to engineering units. They are at an intermediate level, one step before the conversion to calibrated values. Algorithms provided by the sensor teams are used for these conver-sions. From partially processed data level on, the data tables have a time field computed from raw data, sensor configuration files and timelines.

The calibrated data are obtained by converting engineering units to physical units, with algorithms provided by the sensor teams.

## 4. Data Organization

The data format and conventions follow the PDS4 rules and the ExoMars definitions and conventions. The bundle will contain several collections of the types described below.

- A calibration collection containing data and files necessary for the calibration of basic products.

- A context collection, i.e. the list of products comprising various objects, identified within the PDS4 registry, that are specific to the science bundle. These include physical objects such as instruments, spacecraft, and planets and conceptual objects such as missions and PDS nodes.

- Several data collections that contain observational products, generated according to processing level, target, instrument mode, etc.; these are the science data that most users seek. The DREAMS bundle will consist of several data collections, one for each processing level (raw, partially processed, etc.). Within each data level, a further division is by mission phase. Within each mission phase, a further division is by instrument subunit, i.e. by sensor in the case of DREAMS.

- A document collection including the Experiment to Archive Interface Control Document.

- An XML schema collection containing all XML schema files included in or referenced by XML labels in the bundle along with any Schematron files created for validation purposes.

**References**

Bettanini, C., Esposito, F., Debei, S., Molfese, C., RodrÃ­guez, I. A., Colombatti, G., Harri, A. M., Montmessin, F., Wilson, C., Aboudan, A., Abbaki, S., Apestigue, V., Bellucci, G., Berthelier, J. J., Brucato, J. R., Calcutt, S. B., Cortecchia, F., Achille, G. D., Ferri, F., Forget, F., Guizzo, G. P., Friso, E., Genzer, M., Gilbert, P., Haukka, H., Jimenez, J. J., Jimenez, S., Josset, J. L., Karatekin, O., Landis, G., Lorenz, R., Martinez, J., Marty, L., Mennella, V., MÃ¼hlmann, D., Moirin, D., Molinaro, R., Palomba, E., Patel, M., Pommereau, J. P., Popa, C. I., Rafkin, S., Rannou, P., Renno, N. O., Schipani, P., Schmidt, W., Silvestro, S., Simoes, F., Spiga, A., Valero, F., Vazquez, L., Vivat, F., Witasse, O., Mugnuolo, R., Pirrotta, S., & Marchetti, E. 2014, in Metrology for Aerospace (MetroAeroSpace), 2014 IEEE, 167
Esposito, F. 2015, European Planetary Science Congress 2015, Nantes (France), 10, EPSC2015-364